\documentclass[aps,tightlines]{revtex4}

\usepackage{amsfonts}
\usepackage{amsmath}
\usepackage{amssymb}
\usepackage{graphicx}

\begin{document}

\title[Four\_tangle...]{Four-tangle for pure states}
\author{S. Shelly Sharma}
\email{shelly@uel.br}
\affiliation{Depto. de F\'{\i}sica, Universidade Estadual de Londrina, Londrina
86051-990, PR Brazil }
\author{N. K. Sharma}
\email{nsharma@uel.br}
\affiliation{Depto. de Matem\'{a}tica, Universidade Estadual de Londrina, Londrina
86051-990 PR, Brazil }
\thanks{}

\begin{abstract}
An expression for four-tangle is obtained by examining the negativity fonts
present in a four-way partial transpose under local unitary operations. An
alternate derivation of three tangle is also given.
\end{abstract}

\maketitle


Entanglement is inherent to interacting quantum systems and its
quantification is an important question in quantum mechanics. Besides that
entanglement is a physical resource for quantum communication \cite{eker91},
and a key ingredient for quantum computation \cite{jozs03,orus04}.
Negativity \cite{zycz98} of global partial transpose is a widely used
computable measure of free bipartite entanglement. Negativity is based on
Peres-Horodecki NPT criterion \cite{pere96,horo96} and is known to be an
entanglement monotone \cite{vida02}. It was shown in refs. \cite%
{shar07,shar08,shar09} that the global partial transpose of an $N-$qubit
state may be written as a sum of $K-$way partial transposes. In mathematics,
the positive definiteness of matrix is checked by applying Sylvester's
criterion. Our approach is an application of Sylvester's criterion and
Peres-Horodecki NPT criterion aimed at relating the intrinsic eigenvalues
characterizing partial $K-$way transposes of a composite system state
operator to relevant invariants under local unitaries. A well known measure
of purely tripartite entanglement is residual tangle or three-tangle \cite%
{coff00}, obtained by using concurrence \cite{woot98}, which measures
entanglement of bipartite mixed state. In this letter, we examine the
negativity fonts in a three way partial transpose and present a derivation
of three tangle. In addition, an expression for four-tangle is obtained by
examining the\ behavior of negativity fonts present in a four-way partial
transpose under local unitary operations.

Firstly, what is a negativity font? A general N-qubit pure state reads as%
\begin{equation}
\left\vert \Phi ^{ABC......}\right\rangle
=\sum\limits_{i_{1}i_{2}...i_{N}}a_{i_{1}i_{2}...i_{N}}\left\vert
i_{1}i_{2}...i_{N}\right\rangle ,
\end{equation}%
where $\left\vert i_{1}i_{2}...i_{N}\right\rangle $ are the basis vectors
spanning the $2^{N}$ dimensional Hilbert space. The coefficients $%
a_{i_{1}i_{2}...i_{N}}$ are complex numbers with $i_{m}=0$ and $1,$ where $%
m=1,...,N$. Consider the general two qubit state, 
\begin{equation}
\left\vert \Phi ^{AB}\right\rangle =a_{00}\left\vert 00\right\rangle
_{AB}+a_{10}\left\vert 10\right\rangle _{AB}+a_{01}\left\vert
01\right\rangle _{AB}+a_{11}\left\vert 11\right\rangle _{AB},
\end{equation}%
with associated density matrix $\rho ^{AB}$ representing the state operator $%
\widehat{\rho }^{AB}=\left\vert \Phi ^{AB}\right\rangle \left\langle \Phi
^{AB}\right\vert $. The squared global negativity of four by four matrix $%
\left( \rho ^{AB}\right) _{G}^{T_{A}}$ obtained by partially transposing the
state of qubit $A$ in $\rho ^{AB}$, is given by $\left( N_{G}^{A}\right)
^{2}=4\left\vert \det \left[ 
\begin{array}{cc}
a_{00} & a_{01} \\ 
a_{10} & a_{11}%
\end{array}%
\right] \right\vert $. Negativity is invariant with respect to unitaries U$%
^{A}$ and U$^{B}$. The matrix $\nu ^{00}=\left[ 
\begin{array}{cc}
a_{00} & a_{01} \\ 
a_{10} & a_{11}%
\end{array}%
\right] $ represents the single negativity font present in $\left( \rho
^{AB}\right) ^{T_{A}}$. If $\det \nu ^{00}=0$ the state is separable.

Some definitions are in place here. To simplify the notation we denote the
vector $\left\vert i_{1}i_{2}...i_{N}\right\rangle $ by $\left\vert
\prod\limits_{m=1}^{N}i_{m}\right\rangle $ and write density operator for
the state as 
\begin{equation}
\widehat{\rho }=\sum_{\substack{ i_{1}-i_{N},  \\ j_{1}-j_{N}}}\left\langle
\prod\limits_{m=1}^{N}i_{m}\right\vert \widehat{\rho }\left\vert
\prod\limits_{m=1}^{N}j_{m}\right\rangle \left\vert
\prod\limits_{m=1}^{N}i_{m}\right\rangle \left\langle
\prod\limits_{m=1}^{N}j_{m}\right\vert .  \label{1}
\end{equation}%
The matrix elements of global partial transpose of $\widehat{\rho }$ with
respect to qubit $p$ are related to matrix elements of state operator
through 
\begin{equation}
\left\langle \prod\limits_{m=1}^{n}i_{m}\right\vert \widehat{\rho }%
_{G}^{T_{p}}\left\vert \prod\limits_{m=1}^{n}j_{m}\right\rangle
=\left\langle j_{p}\prod\limits_{m=1,m\neq p}^{n}i_{m}\right\vert \widehat{%
\rho }\left\vert i_{p}\prod\limits_{m=1,m\neq p}^{n}j_{m}\right\rangle .
\end{equation}%
To construct $K-$way partial transposes \cite{shar09}, every matrix element $%
\left\langle \prod\limits_{m=1}^{N}i_{m}\right\vert \widehat{\rho }%
\left\vert \prod\limits_{m=1}^{N}j_{m}\right\rangle $ is labelled by a
number $K=\sum\limits_{m=1}^{N}(1-\delta _{i_{m},j_{m}}),$ where $\delta
_{i_{m},j_{m}}=1$ for $i_{m}=j_{m}$, and $\delta _{i_{m},j_{m}}=0$ for $%
i_{m}\neq j_{m}$. The $K-$way partial transpose ($K>2$) of\ $\rho $ with
respect to subsystem $p$ is obtained by selective transposition such that 
\begin{eqnarray}
\left\langle \prod\limits_{m=1}^{N}i_{m}\right\vert \widehat{\rho }%
_{K}^{T_{p}}\left\vert \prod\limits_{m=1}^{N}j_{m}\right\rangle
&=&\left\langle j_{p}\prod\limits_{m=1,m\neq p}^{N}i_{m}\right\vert \widehat{%
\rho }\left\vert i_{p}\prod\limits_{m=1,m\neq p}^{N}j_{m}\right\rangle , 
\notag \\
\quad \text{if}\quad \sum\limits_{m=1}^{N}(1-\delta _{i_{m},j_{m}})
&=&K,\quad \text{and }\quad \delta _{i_{p},j_{p}}=0  \label{ptk1}
\end{eqnarray}%
and%
\begin{eqnarray}
\left\langle \prod\limits_{m=1}^{N}i_{m}\right\vert \widehat{\rho }%
_{K}^{T_{p}}\left\vert \prod\limits_{m=1}^{N}j_{m}\right\rangle
&=&\left\langle \prod\limits_{m=1}^{N}i_{m}\right\vert \widehat{\rho }%
\left\vert \prod\limits_{m=1}^{N}j_{m}\right\rangle ,  \notag \\
\quad \text{if}\quad \sum\limits_{m=1}^{N}(1-\delta _{i_{m},j_{m}}) &\neq &K.
\label{ptk2}
\end{eqnarray}%
while%
\begin{eqnarray}
\left\langle \prod\limits_{m=1}^{N}i_{m}\right\vert \widehat{\rho }%
_{2}^{T_{p}}\left\vert \prod\limits_{m=1}^{N}j_{m}\right\rangle
&=&\left\langle j_{p}\prod\limits_{m=1,m\neq p}^{N}i_{m}\right\vert \widehat{%
\rho }\left\vert i_{p}\prod\limits_{m=1,m\neq p}^{N}j_{m}\right\rangle , 
\notag \\
\quad \text{if}\quad \sum\limits_{m=1}^{N}(1-\delta _{i_{m},j_{m}}) &=&1%
\text{ or }2,\quad \text{and }\quad \delta _{i_{p},j_{p}}=0  \label{pt21}
\end{eqnarray}%
and%
\begin{eqnarray}
\left\langle \prod\limits_{m=1}^{N}i_{m}\right\vert \widehat{\rho }%
_{2}^{T_{p}}\left\vert \prod\limits_{m=1}^{N}j_{m}\right\rangle
&=&\left\langle \prod\limits_{m=1}^{N}i_{m}\right\vert \widehat{\rho }%
\left\vert \prod\limits_{m=1}^{N}j_{m}\right\rangle ,  \notag \\
\quad \text{if}\quad \sum\limits_{m=1}^{N}(1-\delta _{i_{m},j_{m}}) &\neq &1%
\text{ or }2.  \label{pt22}
\end{eqnarray}

Peres PPT separability criterion \cite{pere96} states that the partial
transpose $\widehat{\rho }_{G}^{T_{p}}$ of a separable state is positive.
Sylvester's criterion states that a Hermitian matrix is positive-definite if
and only if the matrix and all of the leading principal minors are positive.
Principal minors are formed by removing row-column pairs from the original
Hermitian matrix. Peres separability criterion translates to examining the
leading principal minors of $\widehat{\rho }_{G}^{T_{p}}$ for positivity.
Entanglement is detected by demonstrating that there exists at least one
principal minor of $\widehat{\rho }_{G}^{T_{p}}$ which is negative. Let us
examine a four by four sub-matrix of $\widehat{\rho }_{G}^{T_{p}}$ in the
space spanned by vectors $\left\vert i_{1}i_{2}...i_{N}\right\rangle $, $%
\left\vert j_{1}j_{2}...j_{N}\right\rangle $, $\left\vert
i_{1}i_{2}...j_{p}...i_{N}\right\rangle $, and $\left\vert
j_{1}j_{2}...i_{p}...j_{N}\right\rangle $. If $\widehat{\rho }$ is a pure
state, the sub-matrix may have a negative eigenvalue $\lambda
^{-}=-\left\vert
a_{i_{1}i_{2}...i_{N}}a_{j_{1}j_{2}...j_{N}}-a_{i_{1}i_{2}...j_{p}...i_{N}}a_{j_{1}j_{2}...i_{p}...j_{N}}\right\vert 
$ as such is a possible negativity font in $\widehat{\rho }_{G}^{T_{p}}$. In
analogy with the two qubit case, a typical negativity font present in $%
\widehat{\rho }_{G}^{T_{p}}$ is represented by the matrix 
\begin{equation*}
\nu =\left[ 
\begin{array}{cc}
a_{i_{1}i_{2}...i_{N}} & a_{j_{1}j_{2}...i_{p}...j_{N}} \\ 
a_{i_{1}i_{2}...j_{p}...i_{N}} & a_{j_{1}j_{2}...j_{N}}%
\end{array}%
\right] .
\end{equation*}%
By shuffling the basis any one of the possible four by four sub-matrices of
the form $\nu $ in $\widehat{\rho }_{G}^{T_{p}}$ can become a leading
principal minor of $\widehat{\rho }_{G}^{T_{p}}$. Sylvester's criterion in
the context of Peres condition implies that all possible leading four by
four sub-matrices $\widehat{\rho }_{G}^{T_{p}}$ are positive for a separable
state. An entangled state has at least one negative four by four sub-matrix
(a negativity font) in $\widehat{\rho }_{G}^{T_{p}}$. Negativity of a given
font is invariant with respect to a unitary transformation on subsystem $p$.
The Global negativity 
\begin{equation}
N_{G}^{p}=\left( \left\Vert \rho _{G}^{T_{p}}\right\Vert _{1}-1\right) ,
\end{equation}%
arising due to all the negativity fonts present in $\widehat{\rho }%
_{G}^{T_{p}}$ measures the entanglement of subsystem $p$ with it's
complement. Here $\left\Vert \widehat{\rho }\right\Vert _{1}$ is the trace
norm of $\widehat{\rho }$.

The $K-$way negativity calculated from $K-$way partial transpose of matrix $%
\rho $ with respect to subsystem $p$, is defined as $N_{K}^{p}=\left(
\left\Vert \rho _{K}^{T_{p}}\right\Vert _{1}-1\right) $. Using the
definition of trace norm and the fact that $tr(\rho _{K}^{T_{p}})=1$, we get 
$N_{K}^{p}=2\sum_{i}\left\vert \lambda _{i}^{K-}\right\vert $, $\lambda
_{i}^{K-}$ being the negative eigenvalues of matrix $\rho _{K}^{T_{p}}$. The 
$K-$way negativity ($2\leq K\leq N)$, defined as the negativity of $K-$way
partial transpose, is determined by the presence or absence of $K-$way
quantum coherences in the composite system. By $K-$way coherences we mean
the type of coherences present in a $K-$qubit GHZ- like state. The
negativity $N_{K}^{p}$ is a measure of all possible types of entanglement
attributed to $K-$ way coherences.

It is straight forward to verify that 
\begin{equation}
\widehat{\rho }_{G}^{T_{p}}=\sum\limits_{K=2}^{N}\widehat{\rho }%
_{K}^{T_{p}}-(N-2)\widehat{\rho }.  \label{3n}
\end{equation}%
By rewriting the global partial transpose as a sum of $K-$way partial
transposes, the negativity fonts are distributed amongst $N-1$ partial
transposes.

\section{Three-tangle}

The state operator for a general three qubit pure state may be rewritten as%
\begin{equation}
\left\vert \Psi ^{ABC}\right\rangle =\left\vert \Phi
_{000}^{ABC}\right\rangle +\left\vert \Phi _{001}^{ABC}\right\rangle ,\quad 
\widehat{\rho }^{ABC}=\left\vert \Psi ^{ABC}\right\rangle \left\langle \Psi
^{ABC}\right\vert ,
\end{equation}%
where%
\begin{equation*}
\left\vert \Phi _{000}^{ABC}\right\rangle =a_{000}\left\vert
000\right\rangle +a_{111}\left\vert 111\right\rangle +a_{100}\left\vert
100\right\rangle +a_{011}\left\vert 011\right\rangle ,
\end{equation*}%
\begin{equation*}
\left\vert \Phi _{001}^{ABC}\right\rangle =a_{001}\left\vert
001\right\rangle +a_{110}\left\vert 110\right\rangle +a_{101}\left\vert
101\right\rangle +a_{010}\left\vert 010\right\rangle .
\end{equation*}%
The global partially transposed matrix $\left( \rho ^{ABC}\right)
_{G}^{T_{A}}$ is related to three-way and two-way partial transposes
obtained from $\rho ^{ABC}$ by selective partial transposition as in Eqs. (%
\ref{ptk1}-\ref{pt22}) through 
\begin{equation}
\left( \rho ^{ABC}\right) _{G}^{T_{A}}=\left( \rho ^{ABC}\right)
_{3}^{T_{A}}+\left( \rho ^{ABC}\right) _{2}^{T_{A}}-\rho ^{ABC}.
\end{equation}%
The three way partial transpose $\left( \rho ^{ABC}\right) _{3}^{T_{A}}$ has
two negativity fonts 
\begin{equation}
\nu ^{000}=\left[ 
\begin{array}{cc}
a_{000} & a_{011} \\ 
a_{100} & a_{111}%
\end{array}%
\right] ,\quad \text{and }\nu ^{001}=\left[ 
\begin{array}{cc}
a_{001} & a_{010} \\ 
a_{101} & a_{110}%
\end{array}%
\right] .
\end{equation}%
The square of negativity of $\left( \rho _{000}^{ABC}\right) _{3}^{T_{A}}$
is found to be $4\left\vert \det \left( \nu ^{000}\right) \right\vert $ and
that of $\left( \rho _{001}^{ABC}\right) _{3}^{T_{A}}$ is $4\left\vert \det
\left( \nu ^{001}\right) \right\vert $. After a\ local unitary $U^{B}=\frac{1%
}{\sqrt{1+\left\vert x\right\vert ^{2}}}\left[ 
\begin{array}{cc}
1 & -x^{\ast } \\ 
x & 1%
\end{array}%
\right] ,$ the state $\left\vert \Psi ^{ABC}\right\rangle $ reads as $%
U^{B}\left\vert \Psi ^{ABC}\right\rangle
=\sum\limits_{i_{1}i_{2}i_{3}}b_{i_{1}i_{2}i_{3}}\left\vert
i_{1}i_{2}i_{3}\right\rangle $. Defining%
\begin{equation*}
P_{B_{0}}^{00}=\det \left[ 
\begin{array}{cc}
a_{000} & a_{001} \\ 
a_{100} & a_{101}%
\end{array}%
\right] ,\quad P_{B_{1}}^{00}=\det \left[ 
\begin{array}{cc}
a_{010} & a_{011} \\ 
a_{110} & a_{111}%
\end{array}%
\right] ,
\end{equation*}%
\begin{equation*}
T^{000}=\det \left[ 
\begin{array}{cc}
a_{000} & a_{011} \\ 
a_{100} & a_{111}%
\end{array}%
\right] ,\quad T^{001}=\det \left[ 
\begin{array}{cc}
a_{001} & a_{010} \\ 
a_{101} & a_{110}%
\end{array}%
\right] ,
\end{equation*}%
and using primed symbols for similar determinants after unitary
transformation, we obtain%
\begin{equation}
\left( T^{000}\right) ^{\prime }=\frac{1}{1+\left\vert x\right\vert ^{2}}%
\left( T^{000}+\left\vert x\right\vert ^{2}T^{001}-x^{\ast
}P_{B_{1}}^{00}+xP_{B_{0}}^{00}\right) ,  \label{UB1}
\end{equation}%
\begin{equation}
\left( T^{001}\right) ^{\prime }=\frac{1}{1+\left\vert x\right\vert ^{2}}%
\left( T^{001}+\left\vert x\right\vert ^{2}T^{000}+x^{\ast
}P_{B_{1}}^{00}-xP_{B_{0}}^{00}\right) ,  \label{UB2}
\end{equation}%
\begin{equation}
\left( P_{B_{0}}^{00}\right) ^{\prime }=\frac{1}{1+\left\vert x\right\vert
^{2}}\left( P_{B_{0}}^{00}+\left( x^{\ast }\right)
^{2}P_{B_{1}}^{00}+x^{\ast }\left( T^{001}-T^{000}\right) \right) ,
\label{UB3}
\end{equation}%
\begin{equation}
\left( P_{B_{1}}^{00}\right) ^{\prime }=\frac{1}{1+\left\vert x\right\vert
^{2}}\left( P_{B_{1}}^{00}+x^{2}P_{B_{0}}^{00}-x\left(
T^{001}-T^{000}\right) \right) .  \label{UB4}
\end{equation}%
It is easily verified that $\left( T^{001}-T^{000}\right) $, $P_{B1}^{00}$
and $P_{B0}^{00}$ are invariant under local unitaries on qubits $A$ and $C$.
Using the relations given in Eqs. (\ref{UB1}-\ref{UB4}) we obtain the three
qubit invariant $T=\left( T^{001}-T^{000}\right) ^{2}-4\left(
P_{B1}^{00}\right) \left( P_{B0}^{00}\right) $, which determines the
three-tangle \cite{coff00} through 
\begin{equation*}
\tau _{3}=4\left\vert \left( T^{001}-T^{000}\right)
^{2}-4P_{B1}^{00}P_{B0}^{00}\right\vert \text{.}
\end{equation*}%
It is further noted that the product $%
T^{001}T^{000}=P_{C_{0}}^{00}P_{C_{1}}^{00}-P_{B1}^{00}P_{B0}^{00}$,\
therefore alternate form of the invariant is $\left\vert \left(
T^{001}+T^{000}\right) ^{2}-4P_{C_{0}}^{00}P_{C_{1}}^{00}\right\vert $.
Starting from negativity fonts in global partial transpose with respect to $%
B $ and $C$ will yield the same invariant.

\section{Four tangle - A measure of genuine four qubit entanglement}

In the case of a general four qubit state 
\begin{equation*}
\left\vert \Psi ^{ABCD}\right\rangle
=\sum\limits_{i_{1}i_{2}i_{3}i_{4}}a_{i_{1}i_{2}i_{3}i_{4}}\left\vert
i_{1}i_{2}i_{3}i_{4}\right\rangle ,
\end{equation*}%
four posible negativity fonts in $\left( \rho ^{ABCD}\right) _{4}^{T_{A}}$
are identified as 
\begin{eqnarray}
\nu ^{0000} &=&\left[ 
\begin{array}{cc}
a_{0000} & a_{0111} \\ 
a_{1000} & a_{1111}%
\end{array}%
\right] ,\quad \nu ^{0001}=\left[ 
\begin{array}{cc}
a_{0001} & a_{0110} \\ 
a_{1001} & a_{1110}%
\end{array}%
\right] ,  \notag \\
\nu ^{0010} &=&\left[ 
\begin{array}{cc}
a_{0010} & a_{0101} \\ 
a_{1010} & a_{1101}%
\end{array}%
\right] ,\quad \nu ^{0011}=\left[ 
\begin{array}{cc}
a_{0011} & a_{0100} \\ 
a_{1011} & a_{1100}%
\end{array}%
\right] .
\end{eqnarray}%
Here qubits $A$ and $B$ are chosen as leading qubits. A unitary on qubit $A$
does not change $\det \left( \nu ^{00i_{3}i_{4}}\right) $.

After applying unitary transformation to qubit $D$ we obtain 
\begin{equation}
U^{D}\left\vert \Psi ^{ABCD}\right\rangle
=\sum\limits_{i_{1}i_{2}i_{3}i_{4}}d_{i_{1}i_{2}i_{3}i_{4}}\left\vert
i_{1}i_{2}i_{3}i_{4}\right\rangle .
\end{equation}%
Defining 
\begin{equation}
F_{00i_{3}i_{4}}=\det \left[ 
\begin{array}{cc}
a_{00i_{3}i_{4}} & a_{01i_{3}+1i_{4}+1} \\ 
a_{10i_{3}i_{4}} & a_{11i_{3}+1i_{4}+1}%
\end{array}%
\right] ,\quad T_{C_{i_{3}}}^{00i_{4}}=\det \left[ 
\begin{array}{cc}
a_{00i_{3}i_{4}} & a_{01i_{3}i_{4}+1} \\ 
a_{10i_{3}i_{4}} & a_{11i_{3}i_{4}+1}%
\end{array}%
\right] ,
\end{equation}%
and using primed symbols for the same quantities calculated from
coefficients $d_{i_{1}i_{2}i_{3}i_{4}}$ , we can verify that 
\begin{equation}
\left( F_{0001}^{\prime }-F_{0000}^{\prime }\right) \pm \left(
F_{0010}^{\prime }-F_{0011}^{\prime }\right) =\left(
F_{0001}-F_{0000}\right) \pm \left( F_{0010}-F_{0011}\right) .  \label{UD}
\end{equation}%
Now application of unitary $U^{C}=\frac{1}{\sqrt{1+\left\vert y\right\vert
^{2}}}\left[ 
\begin{array}{cc}
1 & -y^{\ast } \\ 
y & 1%
\end{array}%
\right] $ yields the state $U^{C}\left\vert \Psi ^{ABCD}\right\rangle
=\sum\limits_{i_{1}i_{2}i_{3}i_{4}}c_{i_{1}i_{2}i_{3}i_{4}}\left\vert
i_{1}i_{2}i_{3}i_{4}\right\rangle ,$ such that 
\begin{eqnarray}
F_{0001}^{\prime \prime }-F_{0000}^{\prime \prime } &=&\frac{1}{\sqrt{%
1+\left\vert y\right\vert ^{2}}}\left[ \left( F_{0001}-F_{0000}\right)
+\left\vert y\right\vert ^{2}\left( F_{0010}-F_{0011}\right) \right.  \notag
\\
&&\left. +y^{\ast }\left( T_{C_{1}}^{000}-T_{C_{1}}^{001}\right) -y\left(
T_{C_{0}}^{000}-T_{C_{0}}^{001}\right) \right] ,
\end{eqnarray}%
\begin{eqnarray}
F_{0010}^{\prime \prime }-F_{0011}^{\prime \prime } &=&\frac{1}{\sqrt{%
1+\left\vert y\right\vert ^{2}}}\left[ \left( F_{0010}-F_{0011}\right)
+\left\vert y\right\vert ^{2}\left( F_{0001}-F_{0000}\right) \right.  \notag
\\
&&\left. -y^{\ast }\left( T_{C_{1}}^{000}-T_{C_{1}}^{001}\right) +y\left(
T_{C_{0}}^{000}-T_{C_{0}}^{001}\right) \right] ,
\end{eqnarray}%
yielding 
\begin{equation}
\left( F_{0001}^{\prime \prime }-F_{0000}^{\prime \prime }\right) +\left(
F_{0010}^{\prime \prime }-F_{0011}^{\prime \prime }\right) =\left(
F_{0001}-F_{0000}\right) +\left( F_{0010}-F_{0011}\right) \text{.}
\label{UC}
\end{equation}

To verify how four-way negativity fonts transform under unitary $U^{B}=\frac{%
1}{\sqrt{1+\left\vert z\right\vert ^{2}}}\left[ 
\begin{array}{cc}
1 & -z^{\ast } \\ 
z & 1%
\end{array}%
\right] $, we write $U^{B}\left\vert \Psi ^{ABCD}\right\rangle
=\sum\limits_{i_{1}i_{2}i_{3}i_{4}}b_{i_{1}i_{2}i_{3}i_{4}}\left\vert
i_{1}i_{2}i_{3}i_{4}\right\rangle $. Defining the determinant of three-way
negativity font for qubits ACD in $\left\vert \Psi ^{ABCD}\right\rangle $ as%
\begin{equation}
T_{B_{i_{2}}}^{00i_{4}}=\det \left[ 
\begin{array}{cc}
a_{0i_{2}0i_{4}} & a_{0i_{2}1i_{4}+1} \\ 
a_{1i_{2}0i_{4}} & a_{1i_{2}1i_{4}+1}%
\end{array}%
\right] ,
\end{equation}%
it is found that%
\begin{equation}
F_{0000}^{\prime \prime \prime }=\frac{1}{\sqrt{1+\left\vert z\right\vert
^{2}}}\left[ F_{0000}+\left\vert z\right\vert ^{2}F_{0011}-z^{\ast
}T_{B_{1}}^{000}+zT_{B_{0}}^{000}\right] ,
\end{equation}%
\begin{equation}
F_{0001}^{\prime \prime \prime }=\frac{1}{\sqrt{1+\left\vert z\right\vert
^{2}}}\left[ F_{0001}+\left\vert z\right\vert ^{2}F_{0010}-z^{\ast
}T_{B_{1}}^{001}+zT_{B_{0}}^{001}\right] ,
\end{equation}%
\begin{equation}
F_{0011}^{\prime \prime \prime }=\frac{1}{\sqrt{1+\left\vert z\right\vert
^{2}}}\left[ F_{0011}+\left\vert z\right\vert ^{2}F_{0000}+z^{\ast
}T_{B_{1}}^{000}-zT_{B_{0}}^{000}\right] 
\end{equation}%
\begin{equation}
F_{0010}^{\prime \prime \prime }==\frac{1}{\sqrt{1+\left\vert z\right\vert
^{2}}}\left[ F_{0010}+\left\vert z\right\vert ^{2}F_{0001}+z^{\ast
}T_{B_{1}}^{001}-zT_{B_{0}}^{001}\right] ,
\end{equation}%
resulting in 
\begin{equation}
\left( F_{0001}^{\prime \prime \prime }+F_{0010}^{\prime \prime \prime
}\right) \pm \left( F_{0000}^{\prime \prime \prime }+F_{0011}^{\prime \prime
\prime }\right) =\left( F_{0001}+F_{0010}\right) \pm \left(
F_{0000}+F_{0011}\right) \text{.}  \label{UB}
\end{equation}%
Combining the results of Eqs. (\ref{UD}), (\ref{UC}), and (\ref{UB}) with
the fact that a unitary on qubit $A$ does not change $\det \left( \nu
^{00i_{3}i_{4}}\right) $ the four qubit invariant is found to be $%
F^{ABCD}=\left( F_{0001}-F_{0000}\right) +\left( F_{0010}-F_{0011}\right) $.
The measure of four qubit\ entanglement purely due to four-way correlations
is defined as 
\begin{equation}
\tau _{4}=4\left\vert \left[ \left( F_{0001}-F_{0000}\right) +\left(
F_{0010}-F_{0011}\right) \right] ^{2}\right\vert .
\end{equation}%
We choose to call it four tangle since it measures four qubit GHZ state like
entanglement in the same sense as Wootters's three tangle does for three
qubits.

To summarize, an expression for four-tangle, a collective property of four
qubits,\ has been obtained by using simple mathematics. Four tangle can be
calculated from the coefficients in a general four qubit state without
calculating the canonical state. The approach presented can be recursively
applied to obtain other invariants as well as meaningful invariants in
larger systems. The method may also elucidate the redistribution of quantum
correlations in mixed states and help in obtaining eventually the
entanglement measures for such states.

\end{document}